\documentclass[prl,superscriptaddress,twocolumn,showpacs,amsmath]{revtex4}

\usepackage{bm}
\usepackage{graphicx}

\newcommand{\bra}[1]{\langle #1 | \,}
\newcommand{\ket}[1]{\, | #1 \rangle}

\newcommand{\expv}[1]{\langle #1 \rangle}
\newcommand{\la}{\lambda}

\newcommand{\om}{\omega}
\newcommand{\Om}{\Omega}
\newcommand{\ga}{\gamma}
\newcommand{\Ga}{\Gamma}
\newcommand{\de}{\delta}
\newcommand{\De}{\Delta}
\newcommand{\ka}{\kappa}
\newcommand{\eps}{\epsilon}
\newcommand{\veps}{\varepsilon}
\newcommand{\lra}{\leftrightarrow}

\begin{document}

\title{Reversible state transfer between 
superconducting qubits and atomic ensembles}

\author{David Petrosyan}
\affiliation{Department of Chemical Physics, Weizmann Institute of Science,
Rehovot 76100, Israel}
\affiliation{Institute of Electronic Structure \& Laser, FORTH, 71110
Heraklion, Crete, Greece}
\author{Guy Bensky}
\author{Gershon Kurizki}
\affiliation{Department of Chemical Physics, Weizmann Institute of Science,
Rehovot 76100, Israel}
\author{Igor Mazets}
\affiliation{Atominstitut der \"Osterreichischen Universit\"aten,
TU-Wien, A-1020 Vienna, Austria}
\affiliation{A.F. Ioffe Physico-Technical Institute, 
194021 St.Petersburg, Russia}
\author{Johannes Majer}
\author{J\"org Schmiedmayer}
\affiliation{Atominstitut der \"Osterreichischen Universit\"aten,
TU-Wien, A-1020 Vienna, Austria}

\date{\today}

\begin{abstract}
We examine the possibility of coherent, reversible information transfer 
between solid-state superconducting qubits and ensembles of ultra-cold atoms.
Strong coupling between these systems is mediated by a microwave 
transmission line resonator that interacts near-resonantly with the 
atoms via their optically excited Rydberg states. The solid-state qubits 
can then be used to implement rapid quantum logic gates, while collective 
metastable states of the atoms can be employed for long-term storage and 
optical read-out of quantum information. 
\end{abstract}


\maketitle

%\paragraph*{Motivation.}

Solid state superconducting (SC) qubits \cite{DevMart,YouNori,ssqRev} 
are promising candidates for implementing quantum information (QI) 
processing in a scalable way \cite{QCcomp}. Very fast and efficient 
quantum logic gates can be performed by SC qubits without significant
loss of coherence. However, dephasing and decoherence hinders the 
long-term storage of QI in such qubits. It would therefore be 
desirable to reversibly transfer the QI from the rather fragile 
qubits to a longer-lived system for storage and retrieval purposes 
\cite{RDDLSZ-KTKM,VZKMRS}.

Appealing candidates for such storage are ground electronic (hyperfine) 
states of ultra-cold (UC) atoms having very long coherence times \cite{THSHR}.
QI can then be stored for many seconds as a collective spin excitation 
of an atomic ensemble. Using stimulated Raman techniques, such as 
electromagnetically induced transparency \cite{EITrev}, this QI can 
be optically read out by mapping the collective atomic state onto the 
generated photon, acting as traveling qubit \cite{sphsSPRS}, whose 
detection is possible with quantum efficiency approaching unity.

Here we put forward a proposal for reversible transfer of QI between 
SC qubits and UC atomic ensembles. In our scheme, the coupling between 
these systems is mediated by a near-resonant microwave transmission line
resonator \cite{BHWGS,MCGKJS-SPS,RJSSMG} employing optical excitations 
of atomic Rydberg states \cite{SvdWCL,DPMF}. This hybrid scheme,
despite considerable challenges, is predicted to be feasible and 
allow high-fidelity QI processing.

%\paragraph*{The system.}

Our solid state qubit is represented by a SC Cooper pair box coupled to 
a SC electrode via two tunnel junctions at the rate $E_J$ (charge qubit) 
in the SQUID configuration \cite{ssqRev}. At the charge 
degeneracy point, the energy separation $\hbar \om_{10} = 
2E_J \cos(\pi \Phi /\Phi_0)$ between the qubit states 
$\ket{0}$ and $\ket{1}$ can be dynamically controlled via an external 
magnetic field $B_{\perp}$ that induces flux $\Phi = A B_{\perp}$ through 
the SQUID area $A$ ($\Phi_0 = h c/2 e$ is the flux quantum). Typical 
values for $E_J/\hbar$ are the microwave range ($10-20\,$GHz). 
The dipole moment $\wp_{01} $ for the transition $\ket{0} \lra \ket{1}$ 
is typically very large, $\wp_{01} \simeq 10^{4} \, a_0 e$. Using resonant
microwave fields, one can then perform many fast quantum logic gates 
within the qubit dephasing time $1/\ga_q \gtrsim 1 \, \mu$s \cite{transmon}.

\begin{figure}[b]
\includegraphics[width=0.48\textwidth]{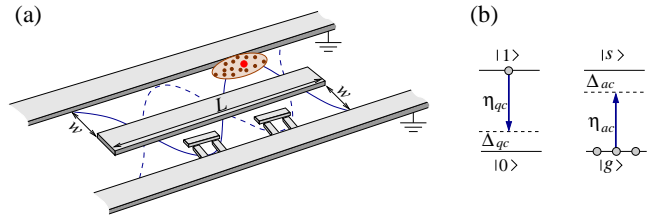}
\caption{(a) CPW cavity with strip-line length $L$ and 
electrode distance $w$. SC qubits are placed at the 
antinodes of the standing wave field and ensembles 
of UC atoms are trapped near the CPW surface.
(b) SC qubit (left) and ensemble qubit (right) can couple 
to a common mode of the CPW cavity.} 
\label{fig:slr}
\end{figure}

Charge qubits can be embedded in near-resonant SC transmission line 
resonators, such as a coplanar waveguide (CPW) cavity of 
Fig.~\ref{fig:slr}(a) \cite{BHWGS,MCGKJS-SPS,RJSSMG}, having 
high quality factor $Q \approx 10^6$. The tight confinement of
the cavity field in a small volume (see below) yields very large 
field per photon $\veps_c$ and strong coupling (vacuum Rabi frequency) 
$\eta_{qc} = (\wp_{01} / \hbar) \veps_c u(\mathbf{r}) 
\sim 2\pi \times 50\,$MHz between the cavity field and the qubit 
located at position $\mathbf{r}$ near the field antinode where 
the cavity mode function $u(\mathbf{r}) \lesssim 1$. In the frame 
rotating with the cavity field frequency $\om_c$, the Hamiltonian 
has the form
\begin{equation}
H_{qc} = \hbar \Delta_{qc} \hat{\sigma}^+ \hat{\sigma}^- 
- \hbar \eta_{qc} (\hat{\sigma}^+ \hat{c} + \hat{c}^{\dag} \hat{\sigma}^- ) , 
\label{Hamqc} 
\end{equation}
where $\Delta_{qc} = \om_{10} - \om_c$ is the externally controlled 
(via $B_{\perp}$) detuning, $\hat{\sigma}^-$ ($\hat{\sigma}^+$) is the 
qubit lowering (rising) operator, and $\hat{c}$ ($\hat{c}^{\dag}$) is 
the cavity photon annihilation (creation) operator.     

One can incorporate many SC qubits in the same CPW cavity, each 
qubit located near the cavity field antinode [Fig.~\ref{fig:slr}(a)]. 
The cavity can then mediate long-range controlled interactions between 
pairs of resonant qubits \cite{BHWGS,MCGKJS-SPS,RJSSMG}, realizing, 
e.g., the two-qubit $\sqrt{\textsc{swap}}$ gate, which together 
with the single qubit rotations form the universal set of logic gates 
in such a quantum computer. However, due to rapid dephasing and relaxation,
neither SC qubits nor the cavity mode can carry out reliable long term 
storage of QI. In what follows, we show that this task can be accomplished
by reversibly transferring the QI to the ground hyperfine states of UC 
atoms trapped near the surface of an integrated atom chip, incorporating 
the CPW cavity and SC qubits \cite{atchip}.

We envision a small trapping volume containing $N \simeq 10^6$ atoms
of $^{87}$Rb with the ground state hyperfine splitting 
$\om_{sg} /2 \pi = 6.83\,$GHz between $\ket{F=1} \equiv \ket{g}$ 
and $\ket{F=2} \equiv \ket{s}$. Let us choose the frequency of the 
CPW cavity to be near-resonant with that of the atomic transition 
$\ket{g} \lra \ket{s}$ [Fig.~\ref{fig:slr}(b)]. When the atomic 
ensemble is near the field antinode, with the spacial dimension 
of the cloud being small compared to the mode wavelength, all the 
atoms couple symmetrically to the cavity field. 
The corresponding Hamiltonian can be expressed as
\begin{equation}
H_{ac} = \hbar \Delta_{ac} \hat{s}^{\dag} \hat{s} 
+ \hbar \eta_{ac} (\hat{s}^{\dag} \hat{g} \, \hat{c} 
+ \hat{c}^{\dag} \hat{g}^{\dag} \hat{s}) \label{Hamac} ,
\end{equation}
where $\Delta_{ac} = \om_{sg} - \om_c$ is the detuning and  
$\eta_{ac} = i (\wp_{sg}/\hbar) \veps_c u(\mathbf{r})$ is the 
coupling rate between the cavity field and a (single) atom at 
position $\mathbf{r}$. Since $\ket{F=1} \lra \ket{F=2}$ is a 
magnetic dipole transition, the corresponding matrix element  
is small, $\wp_{sg} \simeq \frac{1}{2}  \alpha a_0 e$ with $\alpha =1/137$,
which yields $\eta_{ac} \sim 2 \pi \times 20\:$Hz 
[for $u(\mathbf{r}) \lesssim 1$]. The operators $\hat{g}$ ($\hat{g}^{\dag}$)
and $\hat{s}$ ($\hat{s}^{\dag}$) annihilate (create) an atom in the 
corresponding state $\ket{g}$ and $\ket{s}$; these essentially bosonic
operators live in a space of completely symmetrized states $\ket{n_g, n_s}$
with $n_g$ atoms in state $\ket{g}$ and $n_s$ atoms in state 
$\ket{s}$, while $n_g + n_s = N$. 

Apparently \cite{VZKMRS}, the most direct approach to state transfer 
from the SC qubit to the atomic ensemble would be to prepare all
the atoms in state $\ket{g}$, the cavity field in vacuum $\ket{0_c}$,
and choose the frequencies of the cavity mode and the atomic hyperfine 
transition to be the same, $\De_{ac}= 0$. Then, by tuning the SC
qubit frequency to resonance with the cavity, $\De_{qc}= 0$, during 
time $\tau_{qc}$ such that $2 \eta_{qc} \tau_{qc} = \pi$, an arbitrary quantum 
state $\ket{\psi} = \alpha \ket{0} + \beta \ket{1}$ will be transferred 
from the SC qubit to the CPW cavity field [cf. Eq.~(\ref{Hamqc})]. 
Next, it follows from Eq.~(\ref{Hamac}) that the collective coupling 
rate of the cavity field and the atomic ensemble via the transition 
$\ket{n_g= N,n_s=0;1_c} \to \ket{n_g=N-1,n_s=1;0_c}$ is given by
$\sqrt{N} \eta_{ac} \sim 2 \pi \times 20\,$KHz. Thus, during time 
$\tau_{sg} = \pi/(2\sqrt{N} \eta_{ac}) \sim 12 \, \mu$s the cavity photon 
will be absorbed by the atoms and we will have achieved our goal.
The time $\tau_{sg}$ is, however, comparable to the photon lifetime 
in the CPW cavity, $\ka^{-1} = Q/\om_c  \sim 20 \, \mu$s. 
Thus the photon will be lost with high probability 
$P_{\textrm{loss}} \approx \ka \tau_{sg} \sim 0.5$ before being
coherently absorbed by the atoms. It is therefore necessary
to improve the CPW cavity by increasing its quality factor 
$Q$ and thereby decreasing the photon decay rate $\ka$.

In an alternative setup, the SC qubit and the atoms are 
tuned to be resonant with each other, $\De_{qc,ac} \simeq \De $, 
but detuned from the cavity mode frequency. For large detuning 
$\De \gg \eta_{qc}$, the adiabatic elimination of the cavity 
mode yields an effective photon decay rate 
$\kappa_{\textrm{eff}} = \kappa \eta^2_{qc} /\De^2$, while 
the corresponding second-order interaction Hamiltonian,
$V_{qa}^{(2)} = \hbar \eta_{\textrm{eff}} 
(\hat{s}^{\dag} \hat{g} \, \hat{\sigma}^- 
+ \hat{\sigma}^+ \hat{g}^{\dag} \hat{s} )$, with  
$\eta_{\textrm{eff}} = \eta_{qc} \eta_{ac}/\De$, describes an effective 
swap of an excitation between the SC qubit and atomic ensemble with the 
rate $\sqrt{N} \eta_{\textrm{eff}}$, mediated by virtual photon exchange 
in the cavity \cite{Starkq}. Thus the effective coupling is reduced by
a factor of $\De/\eta_{qc}$, while the decoherence rate is reduced by 
a factor of $(\De/\eta_{qc})^2$. For $\De = 10 \eta_{qc}$ we then have 
$\sqrt{N} \eta_{\textrm{eff}}  \sim 2 \pi \times 2\,$KHz 
while $\kappa_{\textrm{eff}} \simeq  2 \pi \times 100\,$Hz, which yields 
a low probability of photon decay during the excitation swap, 
$P_{\textrm{loss}} \approx \kappa_{\textrm{eff}} \pi/(2 \sqrt{N} \eta_{\textrm{eff}}) 
= 0.08$.

Recall, however, that the SC qubit dephasing $\ga_q \lesssim 1 \:$MHz 
is much larger than $\sqrt{N} \eta_{\textrm{eff}}$, hampering the foregoing 
scheme. In turn, $\ga_q$ is much smaller than the coupling $\eta_{qc}$, 
hence this decoherence does not pose a problem during the resonant 
excitation exchange between the SC qubit and CPW cavity. This rapid 
stage may therefore be accomplished with high fidelity, as opposed to the 
much slower stage of excitation transfer from the cavity to the atoms 
with weak magnetic dipole transition. 

\begin{figure}[t]
\includegraphics[width=0.42\textwidth]{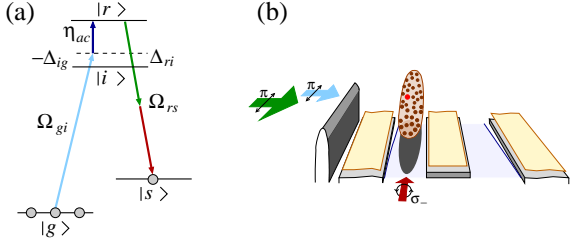}
\caption{(a) Atomic Rydberg states $\ket{i}$ and $\ket{r}$ and 
relevant couplings for the excitation transfer from the CPW cavity 
to the atomic storage state $\ket{s}$.
(b) Propagation geometry of the corresponding optical fields 
and the beam-stop and metallic mirrors on top of the SC electrodes 
of the CPW cavity.} 
\label{fig:ryd}
\end{figure}

Very strong atom--cavity field coupling can be achieved at microwave 
frequencies for electric-dipole transitions between highly-excited 
Rydberg states \cite{RydAtoms,RBHrev}. Let us therefore select a pair
of Rydberg states $\ket{i}$ and $\ket{r}$ such that the frequency 
$\om_{ri}$ of transition $\ket{i} \lra \ket{r}$ is close to the 
cavity mode frequency $\om_c$, the corresponding detuning being 
$\De_{ri} = \om_{ri} - \om_c$. We envision a level scheme sketched 
in Fig. \ref{fig:ryd}(a), where the transition $\ket{g} \lra \ket{i}$ 
is driven by an external optical field with Rabi frequency $\Om_{gi}$ 
and detunings $\De_{ig}$. The Hamiltonian reads 
\begin{eqnarray}
H_{ac} &=& \hbar \Delta_{ig} \hat{i}^{\dag} \hat{i} + 
\hbar (\Delta_{ig} + \Delta_{ri}) \hat{r}^{\dag} \hat{r} \nonumber \\
& & - \hbar ( \Om_{gi} \, \hat{i}^{\dag} \hat{g} 
+ \eta_{ac} \, \hat{r}^{\dag} \hat{i} \, \hat{c} + \mathrm{H.c.}) , 
\label{HamaRc} 
\end{eqnarray} 
where $\eta_{ac} = (\wp_{ir}/\hbar) \veps_c  u(\mathbf{r})$ is the 
atom-cavity field coupling rate, $\wp_{ir}$ being the corresponding
dipole matrix element, while operators $\hat{i}$ ($\hat{i}^{\dag}$) 
and $\hat{r}$ ($\hat{r}^{\dag}$) annihilate (create) an atom in  
state $\ket{i}$ and $\ket{r}$, respectively. 

We set the detunings as $\De_{ri} \simeq - \De_{ig} = \De$. 
Then, given a photon in the cavity, the external field $\Om_{gi}$ 
and the cavity field induce a two-photon transition from the 
ground state $\ket{g}$ to the Rydberg state $\ket{r}$ via 
non-resonant intermediate Rydberg state $\ket{i}$. If 
$\De \gg \eta_{ac}, \sqrt{N} \Om_{gi}$, state $\ket{i}$ 
is never populated, and we obtain an effective interaction
Hamiltonian $V_{ac}^{(2)} = \hbar \eta_{\textrm{eff}} 
(\hat{r}^{\dag} \hat{g} \, \hat{c} + \hat{c}^{\dag} \hat{g}^{\dag} \hat{r})$,
with $\eta_{\textrm{eff}} = \Om_{gi} \eta_{ac}/\De$ \cite{Starka}.
Thus, starting from the initial state of the system 
$\ket{n_g= N,n_{i,r,s}=0;1_c}$, by pulsing $\Om_{gi}$ for time 
$\tau_{gr} = \pi /(2 \sqrt{N} \eta_{\textrm{eff}})$, the cavity 
photon will be coherently absorbed and a single atom from the 
ensemble will be excited to the Rydberg state $\ket{r}$. 
Next, another (bichromatic) external field with Rabi frequency 
$\Om_{rs}$ pulsed for a time $\tau_{rs} = \pi/(2 \Om_{rs})$ can resonantly 
transfer the single collective Rydberg excitation of the atomic ensemble 
to the storage state $\ket{s}$: this process is described by 
$H_{rs} = - \hbar \Om_{rs} \, \hat{s}^{\dag} \hat{r} + \mathrm{H.c.}$ 
At a later time, when required, the reverse process can add a single 
photon in the cavity while all the atoms will end up in state 
$\ket{g}$. This single photonic excitation can then be quickly 
transferred to the SC qubit, as described above. 

Before proceeding, we survey the relevant experimental parameters.
An elongated trapping volume $V_a \sim d \times d \times l$, with 
$d \simeq 5\, \mu$m and $l \simeq 1\,$mm, contains $N \simeq 10^6$ 
atoms at density $\rho_a \sim 4 \times 10^{13}\:$cm$^{-3}$. 
The atomic lower states $\ket{g}$ and $\ket{s}$ correspond to the 
$\ket{F=1,M_F=-1}$ and $\ket{F=2,M_F=1}$ sublevels of the ground 
electronic state $5s_{1/2}$ of $^{87}$Rb. We choose the Rydberg 
states $\ket{i} \equiv \ket{n p_{1/2}, F=2,M_F =-1}$ and  
$\ket{r} \equiv \ket{(n+1) s_{1/2}, F=1,M_F =0}$ with $n=68$ the
principal quantum number. The quantum defects for the $s_{1/2}$ 
and $p_{1/2}$ Rydberg states of Rb are $\de_s = 3.131$ and 
$\de_p = 2.6545$ \cite{RydAtoms}, with which the corresponding 
transition frequency is $\om_{ri} = 2 \pi \times 12.2\,$GHz. 
Calculation of the relevant transition dipole matrix element 
involving the radial and angular parts gives $\wp_{ir} \simeq 1520 a_0 e$.

With the strip-line length $L \simeq 1\:$cm and effective dielectric 
constant $\eps_r \sim 6$, the frequency of the $m$th standing-wave 
mode of the cavity is $\om_c = \pi m c/L \sqrt{\eps_r}$ 
[Fig. \ref{fig:slr}(a)]. The grounded SC electrodes at distance 
$w \simeq 10\;\mu$m confine the cavity field within the effective 
volume $V_c = \int d^3 r |u(\mathbf{r})|^2 \simeq \frac{\pi}{2} w^2 L$ 
yielding $\veps_c = \sqrt{\hbar \om_c/2 \eps_0 V_c} \gtrsim 0.5 \:$V/m.
Taking the full-wavelength ($m=2$) linearly polarized cavity mode 
with $\om_c/2 \pi \simeq 12.16\,$GHz, we estimate \cite{VZKMRS} that 
at the position of atomic cloud about $10\, \mu$m above the CPW surface
the mode function $u(\mathbf{r}) \simeq e^{-1}$ which yields the vacuum 
Rabi frequency $\eta_{ac} = (\wp_{ir}/\hbar) \veps_c  u(\mathbf{r}) 
\simeq 2 \pi \times 3.85\,$MHz and appropriately large detuning 
$\De_{ri} \simeq 10 \eta_{ac}$. 

The transition $\ket{g} \to \ket{i}$ is driven by linearly 
$\pi$-polarized UV field with wavelength $\la_{ig} \simeq 298 \,$nm 
and detuning $\De_{ig} = - \De_{ri}$. To optimize the transition rate, 
its Rabi frequency is chosen as $\sqrt{N} \Om_{gi} \simeq \eta_{ac}$, 
with which the transfer time is $\tau_{gr} \simeq 0.65\,\mu$s. 
The required UV field intensity at the atomic cloud is 
$I_{gi} = 0.46\: \mathrm{W}\, \mathrm{cm}^{-2}$. 
Next, $\ket{r} \to \ket{s}$ is a two-photon transition 
via non-resonant intermediate state $\ket{5p_{1/2},F=2,M_F=0} = \ket{e}$.
The wavelengths are $\la_{re} \simeq 476\,$nm (linearly $\pi$-polarized field)
and $\la_{es} \simeq 795\,$nm (circularly $\sigma_-$-polarized field).
With the corresponding intensities 
$I_{re} = 440 \: \mathrm{W}\, \mathrm{cm}^{-2}$ and 
$I_{es} = 2.25 \: \mathrm{mW}\, \mathrm{cm}^{-2}$ and intermediate 
detuning $\De_{es} = 2 \pi \times 25\,$MHz, the two-photon Rabi 
frequency is $\Om_{rs} = 2 \pi \times 250\,$KHz leading to the transfer 
time of $\tau_{rs} \simeq 1\,\mu$s. Note that $\tau_{gr}$ and $\tau_{rs}$ 
are short compared to the lifetimes of cavity photon 
$1/\kappa \sim 10\,\mu$s and Rydberg states 
$1/\Ga_R \sim 100\,\mu$s \cite{RydAtoms}.

\begin{figure}[t]
\includegraphics[width=0.45\textwidth]{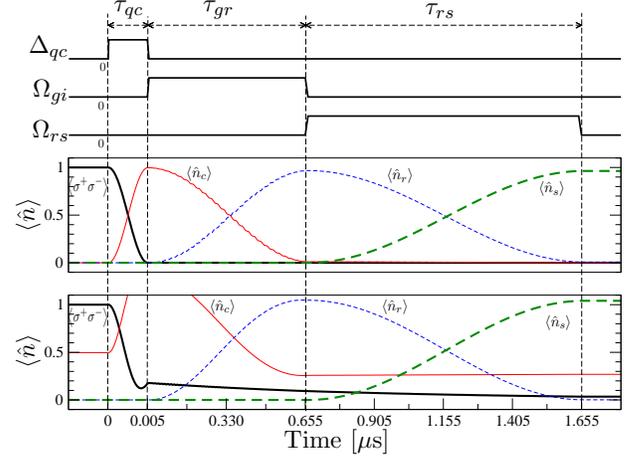}
\caption{Numerical simulations of the three-step excitation 
transfer from the SC qubit to the UC atomic ensemble. 
Top panel illustrates the sequence of ($\pi$-)pulses: 
first the SQ qubit is brought to resonance
with the CPW cavity, by pulsing $\De_{qc}(t) = 0$ for time $\tau_{qc}$;
next, the $\Om_{gi}(t)$ field is pulsed for time $\tau_{gr}$; 
finally, the $\Om_{rs}(t)$ field is pulsed for time $\tau_{rs}$.
Central and lower panels show the dynamics of occupation numbers
for the qubit exited state $\expv{\hat{\sigma}^+ \hat{\sigma}^- }$,
the cavity field $\expv{\hat{n}_c} \equiv \expv{\hat{c}^{\dag} \hat{c}}$
and the collective atomic states 
$\expv{\hat{n}_r} \equiv \expv{\hat{r}^{\dag} \hat{r}}$
and $\expv{\hat{n}_s} \equiv \expv{\hat{s}^{\dag} \hat{s}}$. 
Initially, the mean thermal photon number is $\expv{\hat{n}_c(0)} = 0$
in the central panel, and $\expv{\hat{n}_c(0)} = 0.5$ in the lower panel.} 
\label{fig:proc}
\end{figure}

Employing a master equation approach \cite{QCcomp}, we have simulated 
the dynamics of transfer process in the system with above parameters 
and temperature-dependent initial population $\expv{\hat{n}_c(0)}$ of 
the CPW cavity photon field. Figure~\ref{fig:proc} shows the results 
of numerical integration of the equations for density operator 
$\hat{\rho}(t)$ whose evolution is governed by Hamiltonians 
$H_{qc}$, $H_{ac}$ of (\ref{HamaRc}) and $H_{rs}$. 
As seen, in the case of $\expv{\hat{n}_c(0)} = 0$, the state 
transfer is nearly ideal, with the small final error probability 
$P_{\textrm{err}} \lesssim 0.04$ due to relaxation of the qubit, 
the cavity field and the atoms. However, in the case of finite 
temperature, $\expv{\hat{n}_c(0)} =0.5$, during the transfer, 
as expected, the cavity field and the collective atomic 
state occupation numbers $\expv{\hat{n}_c}$ and $\expv{\hat{n}_{r,s}}$ 
exceed unity, and the resulting error probability 
$P_{\textrm{err}} \simeq 0.3$ is large.

\begin{figure}[t]
\includegraphics[width=0.42\textwidth]{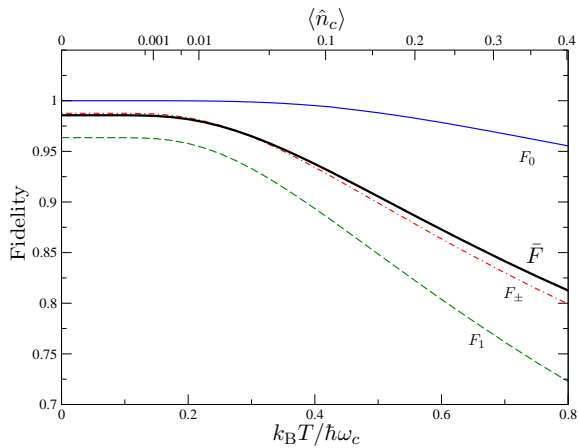}
\caption{Conditional $F_{\psi}$ and mean $\bar{F}$ transfer fidelities 
vs temperature $k_{\textrm{B}} T$ (lower horizontal axis)
or mean thermal photon number $\expv{\hat{n}_c}$ 
(upper horizontal axis). In $F_{\psi}$, $\psi= 0,1,\pm$ correspond to states
$\ket{0}$, $\ket{1}$ and $\frac{1}{\sqrt{2}} [\ket{0} \pm (i) \ket{1}]$.} 
\label{fig:fidelity}
\end{figure}

We can characterize the transfer process for a given initial state
$\ket{\psi}_q$ of the SC qubit by the conditional fidelity 
$F_{\psi} = \mathrm{Tr}(\hat{\rho} \ket{\psi}_{aa}\bra{\psi})$, 
where $\ket{\psi}_a$ denotes the final state stored in the UC 
atomic ensemble for an ideal transfer. The mean transfer fidelity 
$\bar{F}$ is obtained by averaging $F_{\psi}$ over all possible 
$\ket{\psi}$. The dependence of $F_{\psi}$ and $\bar{F}$ on the 
CPW cavity temperature $T$, or the mean thermal photon number 
$\expv{\hat{n}_c} = ( e^{\hbar \om_c / k_{\mathrm{B}} T} -1)^{-1}$, 
is shown in Fig.~\ref{fig:fidelity}. Below 
$k_{\textrm{B}} T/\hbar \om_c \simeq 0.2$, corresponding to 
$\expv{\hat{n}_c} \lesssim 0.01$, the transfer fidelity is fairly 
high, $\bar{F} > 98\%$, but then it quickly degrades due to 
the detrimental effect of even a small number of thermal photons. 
For the above parameters, this critical temperature is 
$T \sim 0.1\:$K necessitating cryogenic conditions. 
We note that the temperature of atomic cloud is much smaller, 
$T_a \lesssim 1\:\mu$K, resulting in the lifetime of atomic 
hyperfine coherence in excess of $1\:$s \cite{THSHR}.
  
The practical realization of our scheme can be hindered by 
absorption of optical fields, driving the atomic transitions, 
by the SC electrodes of the CPW cavity. If not eliminated, 
the photon absorption would break up many Cooper pairs 
and produce quasi-particles resulting in drastic reduction of
the cavity $Q$ factor. Consider a geometry where the strong 
$\pi$-polarized fields propagate in the direction parallel to 
the CPW surface and perpendicular to the electrodes, in front 
of which an opaque barrier of height $\sim 10\: \mu$m serves as 
a beam-stop [Fig.~\ref{fig:ryd}(b)]. Due to the Fresnel diffraction 
on the barrier edge, some light will still reach the shade area behind 
the barrier, but we estimate that the light intensity at the SC electrodes,
each about $10\; \mu$m wide, will be reduced by a factor of $300$. 
However, even after such significant reduction of intensity, the 
residual absorption would remain too large. To completely eliminate 
the absorption, the SC electrodes can be covered by a few $\mu$m 
thick layer of dielectric followed by a thin metallic mirror. 
A moderate reduction of the cavity $Q$ factor up to 10 times 
is tolerable, since the photon lifetime still remains long 
compared to the transfer times $\tau_{qc} + \tau_{gr} \sim 0.7 \:\mu$s, 
but the resulting fidelity will decrease to $\bar{F} \gtrsim 70\%$.

To conclude, we have proposed a promising approach for hybridizing
solid-state and atomic quantum devices, thereby implementing 
efficient quantum state transfer between superconducting charge 
qubits and mesoscopic ultracold atomic ensembles coupled 
to a microwave coplanar waveguide cavity via optically excited
Rydberg transitions. Our scheme is scalable to many superconducting 
qubits, whose merit is very high flipping rates, coupled to atomic 
ensembles serving as reliable storage qubits.  

\begin{acknowledgments}
This research was supported by the EC (MIDAS STREP, FET Open).
D.P. was also supported by the EC Marie Curie RTN EMALI. 
I.M. acknowledges helpful discussions with P. Pillet and D. Comparat.

\end{acknowledgments}


\begin{thebibliography}{99}
\bibitem{Starkq} The second-order level shift of the SC qubit, 
$\eta_{qc}^2/\De$, is assumed to be incorporated in $\De_{qc}$.
\bibitem{Starka} The second-order shift $\eta_{ac}^2/\De$ of the 
atomic Rydberg state $\ket{r}$ is incorporated in $\De_{ri}$.
\end{thebibliography}
\end{document}